# CONFORMAL CORONARY CALCIFICATION VOLUME ESTIMATION WITH CONDITIONAL COVERAGE VIA HISTOGRAM CLUSTERING

*Dr. Olivier Jaubert[1], Mr. Salman Mohammadi[1], Dr. Keith A. Goatman[1], Dr. Shadia S. Mikhael[1], Dr. Conor Bradley[2,3], Mme. Rebecca Hughes[3], Dr. Richard Good[2,3], Dr. John H. Hipwell[1], Dr. Sonia Dahdouh[1]*

[1] Canon Medical Research Europe, [2] University of Glasgow, [3] NHS Golden Jubilee National Hospital

**ABSTRACT**

Incidental detection and quantification of coronary calcium in CT scans could lead to the early introduction of lifesaving clinical interventions. However, over-reporting could negatively affect patient wellbeing and unnecessarily burden the medical system. Therefore, careful considerations should be taken when automatically reporting coronary calcium scores.

A cluster-based conditional conformal prediction framework is proposed to provide score intervals with calibrated coverage from trained segmentation networks without retraining. The proposed method was tuned and used to calibrate predictive intervals for 3D UNet models (deterministic, MCDropout and deep ensemble) reaching similar coverage with better triage metrics compared to conventional conformal prediction.

Meaningful predictive intervals of calcium scores could help triage patients according to the confidence of their risk category prediction.

## 1. INTRODUCTION

Coronary heart disease (CHD) is the prime cause of death worldwide. According to the World Health Organization, in 2019, 32% of all deaths were due to CHD[1]. Coronary artery calcification (CAC) is a highly prognostic biomarker for adverse cardiovascular events[2]. Depending on the risk due to calcium burden, treatments can range from simple lifestyle changes to invasive procedures aiming to revascularize coronary arteries with obstructive atherosclerotic disease. Early discovery in asymptomatic patients could therefore reduce the number of adverse cardiac events whilst detection and quantification in symptomatic patients could guide subsequent investigations including the need for invasive management[3,4]. CAC is often assessed using a dedicated electrocardiogram-gated non-contrast computed tomography (CT) scan. These scans are used to produce an Agatston score, which is the current standard prognostic measure of risk from CAC burden[5]. However, millions of CT scans including the heart in the field of view (FOV) are acquired every year for various indications where visible CAC goes unreported[6]. Most of these scans are not synchronized to the cardiac cycle (un-gated) and contain contrast, compromising the Agatston score calculation. Coronary calcium volume has been shown to be highly correlated with Agatston scores, enabling potential risk stratification from both non-contrast and contrast CT scans[7,8]. There is therefore an opportunity for automated detection and quantification of calcium. This requires algorithms to be robust to a wide range of scan indications, types (e.g. gating and contrast) and image quality (e.g. noise and artefacts).

Automatic predictions "in the wild" may lead to over-reporting, causing unnecessary patient anxiety and additional burden on the medical care system, or under-reporting and a missed opportunity for early intervention. Model uncertainty could help triage the confident cases leaving only the uncertain cases for the clinician to review. In the context of incidental findings and triage, calibrating volume prediction intervals to cover the true volume value as often as possible (i.e., maximize coverage), while preserving helpful clinical information (i.e., tight intervals for confident risk category assessment) could help reduce harmful under- or over-reporting.

This can be performed via conformal prediction (CP), a framework enabling statistically valid calibration of prediction intervals, without retraining of the machine learning model, only assuming exchangeability of data (i.e. that the joint distribution is invariant to finite permutation) [9,10]. Following calibration, the true volume

$Vti$ statistically lies within the interval $PIi$ with the target coverage probability of 1-α:

$$\mathbb{P}(Vti \in PIi) \geq 1 - \alpha$$

Coverage preserved for different conditions (i.e. conditional coverage) is often desirable in medical imaging to reduce the effects of covariate shifts such as differences in image quality, patient cohort, model uncertainty or segmented volume that often occur in real world settings[11].

In this work, we train networks to segment the coronary calcium plaque from CT scans and calibrate predictive intervals of calcium volume using a conditional CP approach. The method introduces a histogram clustering step of the output probabilities. Different clusters reflect differences in the predicted segmentation volumes and uncertainty. The CP calibration is performed to reach target coverage for each cluster and at inference new data is assigned to a cluster to form a cluster-specific conformal predictive interval.

Segmentation models (Deterministic, Monte-Carlo (MC) Dropout and Ensemble) were trained, calibrated with and without cluster-based conditionality, and tested on a dataset of unselected, consecutive CT scans that reflect the incidentals finding setting. Pre-calibrated, conventional, and cluster-based conditional calibrated intervals were compared for the different segmentation models in terms of coverage and confident risk category assignments.

## 2. METHODS

### 2.1. Overview

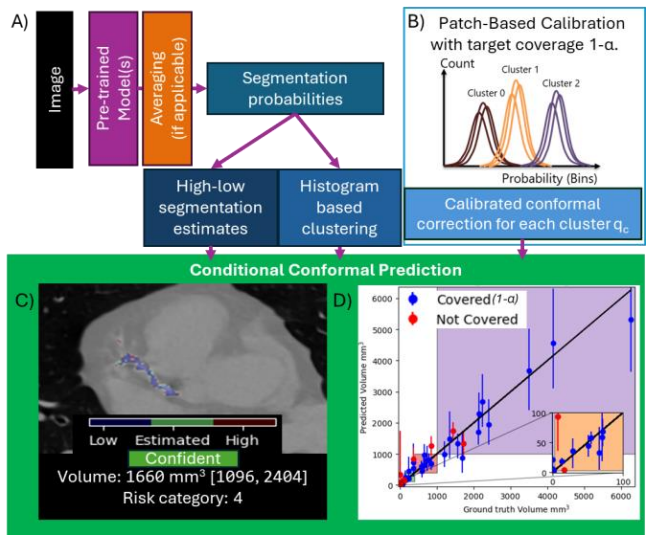

**Fig1. Inference framework for cluster-based conformal coronary calcium prediction. A)** Inference is performed (with an averaging step for ensemble and MC-Dropout methods). **B)** Pre-inference, clusters and corresponding $q_c$ corrections are determined from a calibration dataset. **C)** Conformal intervals of volume are calculated based on the current segmentation probability map and its histogram's nearest cluster. **D)** Low and high-volume estimates are calibrated to obtain the given target coverage (1-α).

An overview of the proposed framework is provided in Fig 1 and 2. From an input CT patch or image, 1) a segmentation model provides a voxel-level probability map of coronary calcium. 2) The volume and pre-calibration interval are estimated via thresholding (Details in 2.2.3). 3) The probability map's histogram is used to identify its closest cluster defined at calibration. 4) The assigned cluster determines which correction (computed at calibration, details in 2.3) to apply to the pre-calibrated intervals to obtain the conditional conformal interval prediction.

```
Inputs: image, model, calibrated correction factors q_i
Inference: prob_map = model(image)
[Li=(prob_map>0.7), Vi=(prob_map>0.5), Hi=(prob_map>0.3)]
histogram= hist(prob_map)
Only at calibration: from all calibration images
clusters=constrained_Kmeans_clustering(calibration_histograms)
for cluster in clusters:
    for image in cluster:
        Si=max(Li-Vti,Vti-Hi) where Vti is the true volume
    q_cluster=quantile(S, [(n+1)(1-α)]/n)
Conformal prediction at testing:
closest_cluster = find_closest_cluster(histogram)
Outputs: [Li-q_closest_cluster, Vi, Hi+q_closest_cluster]
```

**Fig2.** Pseudo-code for cluster-based conformal interval prediction.

### 2.2. Training

#### 2.2.1. Data and pre-processing

295 consecutive, unselected CT scans (containing the entire heart in the FOV) were retrospectively collected at the NHS Golden Jubilee Hospital comprising a wide range of clinical indications, contrast levels and FOVs[12].

Coronary calcium was annotated by one senior (R.G., >10 years of experience) and two junior clinicians (R.H., C.B>2 years of experience). The dataset was split into 142/72/81 for training, calibration, and testing, respectively. To reduce variability in FOVs, TotalSegmentator[13] was used to crop scans to a wide heart FOV input image. An in-house atlas-based segmentation of the heart was used for heart masking to further reduce false positives outside of the heart.

#### 2.2.2. Networks and training parameters

The network was trained on randomly sampled patches of 16x256x256 from the paired heart images and ground truth segmentations, which were then augmented using random rotations and contrast adjustments. We used a PyTorch implementation of the 3D UNet architecture[14] with 32, 64, 128, 256 filters per convolution for the different scales, batch size of 4, a composed Dice and weighted focal loss [14] (w=4450, γ=2), and Adam optimizer (β1 = 0.9, β2 = 0.999, ε = 1e− 8, learning rate = 3x10$^{-4}$).

We additionally investigate the calibration of classically used uncertainty methods: MCDropout[15] and model ensembling[16].

The ensemble of models was trained by varying the random seed for 5 different training runs which affected data augmentation and initialization of the network. For MCDropout, 3D spatial dropout was performed on all but the initial scale convolutions with dropout rate of 0.1 at training and inference similarly to the baseline in [17].

### 2.2.3. Probability maps and pre-calibration intervals

Probabilities were averaged over predictions from each of the 5 models in the ensemble, and over 10 repeated predictions for MCDropout. Pixelwise probability maps were thresholded with threshold $\lambda$ and voxels were summed to obtain the estimated ($Vi, \lambda > 0.5$), low ($Li, \lambda > 0.7$) and high ($Hi, \lambda > 0.3$) pre-calibrated volumes which can be compared to the true volume ($Vti$).

## 2.3. Calibration

### 2.3.1. Conventional Conformal Prediction (cCP) Interval

cCP uses a calibration step which from the pre-calibrated predictive intervals $[Li, Hi]$ computes a correction factor used to obtain the correct target coverage.

To obtain this correction factor, a score function $Si = \max(Li - Vti, Vti - Hi)$ is computed over all the $n$ samples of the calibration set [19]. The quantile $q$ corresponding to the target coverage (with finite element correction): $\frac{[(n+1)(1-\alpha)]}{n}$ of the scores is selected. And conformal predictive intervals are formed as follows: $[Li - q, Hi + q]$.

### 2.3.2. Clustered Conformal Prediction Interval

Although target coverage might be reached globally, a single conformal correction factor $q$ might lead to miscoverage for high or low volumes of calcium. Conditional coverage to obtain correct coverages for different volume uncertainty levels could be beneficial for more confident risk category assignment.

We investigate the use of the probability map's histograms for the clustering of similar cases. The histogram contains non-localized information about coronary calcium uncertainty and volume (i.e. in each bin it contains the count of voxels with given probabilities of being coronary calcium). To remove the influence of the number of background voxels, (up to 99.9% of the voxels in the given sparse segmentation problem), histograms were arbitrarily taken for bin values >0.2 (removing all voxels with probability>80% to be background). This enabled the same clustering for patches (which were used for calibration) and full volume images (for inference).

A constrained K-means clustering algorithm [18] to enforce equal cluster size and equal statistical power for each subsequent calibration was used on the clipped histograms of the probability maps.

Like other clustered approaches [19], conformal calibration as described in Section 2.3.1 is performed to obtain $q_c$ for each cluster. Predictions are assigned to their cluster (according to the minimum Euclidean distance to cluster centers) and the appropriate correction factor $q_c$ is applied.

## 2.4. Experiments

### 2.4.1. Metrics

To assess clinical relevance of the predictive intervals the volumes were converted to estimated Agatston scores by a scaling of 3.13[7] with Agatston score risk categories defined as 0-3: No risk, 3-100: Low, 100-400: Moderate, 400-1000: High and >1000: Very High risk[20]. The boundary of the first risk categories was adjusted from 1 to 3 compared to the original risk classification as classically small calcifications of less than 3 pixels are discarded.

Both empirical coverage and triaging metrics were investigated for different calibrations:

(1) $Coverage = \frac{Ncovered}{NTotal}$

(2) $CA\ rate = \frac{CA}{CA+UA}$; $CE\ rate = \frac{CE}{CE+UE}$

(3) $CAr - CEr = CA\ rate - CE\ rate$

Where $Ncovered$ is the number of points covered (i.e. where $Vti$ lies within $PIi$), CA/CE/UA and UE the number of Confident Accurates/Confident Errors/Uncertain Accurates and Uncertain Errors respectively. A prediction is defined as accurate if the predicted score lies within the correct risk category, an error otherwise. Additionally, a prediction is uncertain if the estimated interval overlaps with both a correct and incorrect risk category and is confident otherwise. Intuitively if $CAr - CEr < 0$, the uncertainty was no better than random guessing at capturing more errors than accurates while $CAr - CEr = 1$ means all errors were predicted as uncertain and all accurates were confident.

### 2.4.2. Patch-Based Calibration Experiments

Calibration experiments were performed on 864 data augmented patches (sampled from 72 patients). For hyperparameter tuning, $N_{runs}$ runs are performed with random half-splitting of the patients into calibration and validation sets. Three calibration experiments were performed to investigate: **1)** different target coverages ($N_{runs}$=5) for both conventional and clustered ($N_{clusters}$=5) approaches, **2)** the optimal number of clusters $N_{clusters}$ ($N_{runs}$=5), and **3)** coverage results for deterministic, MCDropout and ensembled predictions ($N_{runs}$=50).

### 2.4.3. Final Calibration and Full Volume Testing.

Once hyper-parameters were determined, a final calibration step was performed on the full 864 patches of the calibration experiments before testing on the full volume test images.

## 3. RESULTS

### 3.1. Patch Based Calibration

| | Target Coverage | Coverage | CA rate (%) | CE rate (%) | CAr-CEr (%) |
|---|---|---|---|---|---|
| Pre-Calibration | N/A | 0.77 | 70.27 | 45.48 | 24.79 |
| Conventional | 0.75 | 0.79 | 50.38 | 41.94 | 8.44 |
| 5 Clusters | 0.75 | 0.83 | 61.68 | 40.00 | 21.68 |
| Conventional | 0.8 | 0.82 | 23.46 | 30.00 | -6.54 |
| 5 Clusters | 0.8 | 0.85 | 58.38 | 38.06 | 20.31 |
| Conventional | 0.85 | 0.86 | 17.24 | 25.48 | -8.24 |
| **5 Clusters** | **0.85** | **0.87** | **53.41** | **36.13** | **17.28** |
| Conventional | 0.9 | 0.91 | 13.30 | 20.00 | -6.70 |
| 5 Clusters | 0.9 | 0.90 | 22.65 | 29.35 | -6.71 |
| Conventional | 0.95 | 0.94 | 8.00 | 9.68 | -1.68 |
| 5 Clusters | 0.95 | 0.93 | 16.86 | 21.94 | -5.07 |

**Table 1.** Target coverage experiment results averaged over $N_{runs}$=5 runs. Target coverage of 0.85 was selected (in bold)

Target coverage experiment results are presented in Table 1. Mean Coverage Error was 1.6% and 3.3% for conventional and clustered approaches, respectively. However, the clustered predictions had higher CAr-CEr rates over the different target coverages. At target coverage≥ 90%, many uncertain cases arise and CAr-CEr < 0 shows that there is no meaningful risk classification information left. The highest target coverage with CAr-CEr>0 was selected (85%).

| | Coverage (Target=0.85) | CA rate (%) | CE rate (%) | CAr-CEr (%) |
|---|---|---|---|---|
| Pre-Calibration | 0.77 | 70.27 | 45.48 | 24.79 |
| Conventional | 0.86 | 17.24 | 25.48 | -8.24 |
| 2 Clusters | 0.83 | 39.46 | 40.00 | -0.54 |
| **3 Clusters** | **0.86** | **56.86** | **35.48** | **21.38** |
| 4 Clusters | 0.87 | 52.81 | 35.81 | 17.00 |
| 5 Clusters | 0.87 | 53.41 | 36.13 | 17.28 |
| 7 Clusters | 0.87 | 48.43 | 34.84 | 13.59 |
| 10 Clusters | 0.89 | 47.03 | 34.52 | 12.51 |
| 15 Clusters | 0.89 | 38.43 | 30.32 | 8.11 |
| 20 Clusters | 0.89 | 34.32 | 28.06 | 6.26 |

**Table 2.** $N_{clusters}$ experiment results averaged over $N_{runs}$=5 runs (with target coverage of 85%). Highest CAr-CEr values was obtained for 3 clusters (in bold).

Table 2 shows the effect of cluster size $N_{clusters}$ on the results for a target coverage of 85%. Higher numbers of clusters lead to less precise coverage. The best performances, in terms of CAr-CEr, were observed for $N_{clusters}$ =3.
The empirical coverage measured for deterministic, MCDropout and ensembled models over 50 random splits shows that the target coverage of 85% is obtained on average for all models considered for both Conventional and Cluster based calibrations (Fig 3).

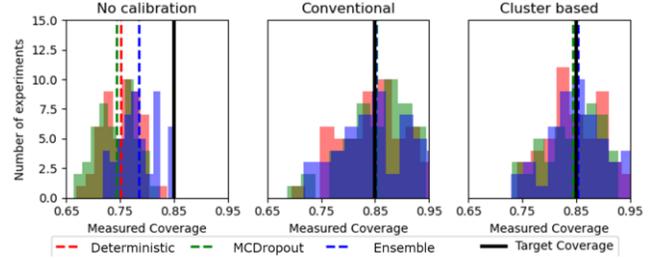

**Fig 3.** Histogram and mean (dotted lines) measured coverage obtained on validation data over 50 random splits for models without calibration, and Conventional and Cluster based Calibration.

### 3.2. Full Volume Predictions

| Models | Calibration Method | Coverage Target=0.85 | CA rate (%) | CE rate (%) | CAr-CEr (%) |
|---|---|---|---|---|---|
| Deterministic | Pre-Calib. | 0.73 | 76.1 | 35.7 | 40.4 |
| | Conventional | 0.80 | 40.3 | 14.3 | 26.0 |
| | 3 Clusters | 0.81 | 65.7 | 14.3 | 51.4 |
| MCDropout | Pre-Calib. | 0.77 | 71.9 | 23.5 | 48.3 |
| | Conventional | 0.81 | 40.6 | 11.8 | 28.9 |
| | 3 Clusters | 0.83 | 60.9 | 11.8 | 49.2 |
| Ensemble | Pre-Calib. | 0.81 | 82.6 | 25.0 | 57.6 |
| | Conventional | 0.85 | 43.5 | 16.7 | 26.8 |
| | 3 Clusters | 0.85 | 78.3 | 16.7 | 61.6 |

**Table 3.** Pre-calibrated, Conventional and Cluster-based calibrated test set results on full image predictions.

Table 3 summarizes test set results (N=81 patients). While pre-calibrated models provide meaningful information with good CAr- CEr, they provide no guarantees on coverage.
When going from patch to full image predictions, CP techniques generally resulted in better coverage. However, cCP led to significantly lower CAr-CEr values.

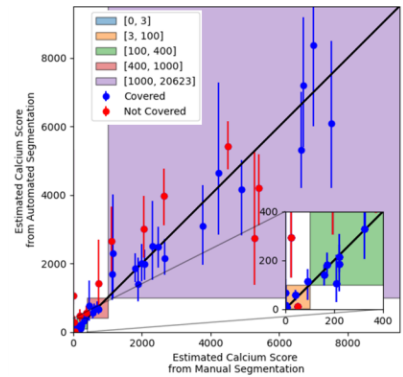

**Fig 4.** Estimated calcium scores and cluster based predicted conformal intervals from the ensemble model obtained on the test data. Risk categories are represented via colored boxes and covered/not covered predictions are represented in blue/red respectively.

Over all models, average test coverage was measured at 0.77, 0.82, and 0.83, and CAr-CEr at 48.8%, 27.2% and 54.1% for pre-calibration, cCP and the proposed method, respectively.

Ensembled results (Fig 4) obtained the highest coverage and CAr-CEr after cluster-based calibration with 54 confident accurate, 15 uncertain accurate, 2 confident erroneous and 10 uncertain erroneous risk category predictions. Fig 4 shows the corresponding predicted scores and intervals.

## 4. DISCUSSION

The proposed cluster-based calibration was compared to the cCP approach for a range of models (Deterministic, MC-Dropout or Ensemble). It enabled targeted coverage of calcium volume with improved triage metrics. With the proposed calibrated ensemble method, reviewing 70% less cases led to only 2.4% errors in risk category predictions.

An estimated Agatston score was derived from the calcium volume [7] as they have been shown to be highly correlated[21]. More sophisticated Agatston estimation methods [4] or risk categories based directly on volume [22] could be used in the future.

We used Euclidean distance for clustering histograms. Using other distances for histogram similarity could be considered, however, a constrained approach with a minimal cluster size for correct calibration is desired for downstream calibration.

We used a retrospective unselected dataset highly relevant for incidental calcium findings. As often seen in medical imaging, the dataset size was limited. Our approach proposed to perform calibration on data augmented patches to increase the statistical power of the calibration in the low data regime. While clustering mitigates the data size issue, moving from patches to full images might mean that the exchangeability hypothesis is no longer respected.

Recently, a weighted conformal prediction approach has been proposed to tackle the non-exchangeability between calibration and test data in medical image segmentation [11]. This however required the training of an additional classifier for each new test dataset (with enough samples for training) which could be a barrier for clinical application.

Nevertheless, when transferring to larger images where a larger relative number of uncertain voxels is expected, our proposed clustering-based CP led to empirical coverages close to target coverage while still preserving high CAr-CEr rate.

## 5. CONCLUSION

In conclusion, the proposed method enabled creation of meaningful predictive intervals with high CAr-CEr rates and coverage outperforming the conventional approach. It required no model re-training using histograms of the output probabilities as a condition for conformal prediction. The method could help reduce the burden on radiologists given both the shortage of specialists and the rise in the number of scans being acquired.

## 6. ACKNOWLEDGMENTS
Six of the authors (O.J, S.M, K.G, S.S.M, J.H, S.D) are employees of Canon Medical Research Europe.

## 7. COMPLIANCE WITH ETHICAL STANDARDS
The study was sponsored by NHS Golden Jubilee, UK and approved by Yorkshire and The Humber – Leeds East Research Ethics Committee (REC Ref 21/YH/0217).